\documentclass[10pt,conference,letterpaper]{IEEEtran}
\IEEEoverridecommandlockouts
\usepackage{cite}
\usepackage{amsmath,amssymb,amsfonts}
\usepackage{textcomp}
\usepackage{xcolor}
\usepackage{hyperref}
\usepackage{CJK}
\def\BibTeX{{\rm B\kern-.05em{\sc i\kern-.025em b}\kern-.08em
    T\kern-.1667em\lower.7ex\hbox{E}\kern-.125emX}}
\usepackage{kotex}

\usepackage{graphicx}
\usepackage{subcaption}
\usepackage{caption}
\usepackage{xurl}

\usepackage{algorithm}
\usepackage{algpseudocode}
\usepackage{multirow}
\usepackage{array, makecell}
\usepackage{booktabs}
\usepackage{tabularx}
\usepackage{xspace}
\usepackage{eso-pic}

\usepackage{enumitem}

\newcommand{\xronos}{\textsc{Xronos}\xspace}

\begin{document}
\bstctlcite{IEEEexample:BSTcontrol}

\title{\xronos: Heterogeneity-Aware Tensor Parallelism for Collaborative LLM Fine-Tuning on Edge CPUs}

\author{%
  \IEEEauthorblockN{
    Wonmi Choi$^{1,*}$,
    Sunjae Park$^{1,*}$,
    Dohyeok Kwon$^{1}$,
    Zhixiong Niu$^{2}$,
    Yeonho Yoo$^{3}$,
    Chuck Yoo$^{1}$,
    Gyeongsik Yang$^{1}$
  }
  \IEEEauthorblockA{
    $^1$Dept. of Computer Science and Engineering, Korea University \quad
    $^2$Microsoft Research Asia \quad
    $^3$Dongguk University
  }
  \thanks{$^*$Equal contribution.}\vspace{-1em}
}

\AddToShipoutPicture* {
  \AtPageLowerLeft {
    \put(0, 40){
      \makebox[\paperwidth][c]{
        \begin{minipage}{\textwidth}
          \centering
          \footnotesize
          \copyright~2026 IEEE. Personal use of this material is permitted. Permission from IEEE must be obtained for all other uses, in any current or future media, including reprinting/republishing this material for advertising or promotional purposes, creating new collective works, for resale or redistribution to servers or lists, or reuse of any copyrighted component of this work in other works. \\
          This paper has been accepted for publication in IEEE MASCOTS 2026. 
        \end{minipage}
      }
    }
  }
}

\maketitle

\begin{abstract}
Collaborative fine-tuning on edge devices adapts large language models to domain-specific data while keeping each device's data local. State-of-the-art (SOTA) collaborative fine-tuning techniques are largely designed for GPU-based edge devices and rely on pipeline parallelism (PP). However, many edge platforms, including IoT gateways, smart-home hubs, and in-vehicle computers, are primarily CPU-based. This paper reports that PP is ineffective on CPU-based edge devices because the same CPU handles both model computation and communication, which causes severe CPU contention. Our analysis shows that this leads to 5.75$\times$ higher computation stall ratios than on GPU devices on average. Tensor parallelism (TP) can alleviate this contention by separating computation and communication, but existing TP techniques assume homogeneous devices. On heterogeneous CPU edge devices, we find that this assumption causes faster workers to remain idle for up to 34\% while waiting for slower devices at synchronization points. To address the limitations, we propose \xronos, a collaborative fine-tuning framework for heterogeneous CPU edge devices. \xronos uses TP as its execution backbone and combines lightweight profiling with heterogeneity-aware tensor partitioning to reduce the straggler bottleneck. Across diverse devices, models, and benchmark tasks, \xronos reduces fine-tuning time by 18\% (TP) to 56\% (PP) and the ratio of device idle time by $\sim$5.9$\times$ over SOTA techniques, while maintaining the accuracy.
\end{abstract}

\begin{IEEEkeywords}
Collaborative fine-tuning, Distributed training modeling, Heterogeneous edge devices, Edge computing
\end{IEEEkeywords}

\section{Introduction}

Collaborative fine-tuning enables multiple edge devices to jointly update a large language model (LLM) without centralizing the raw data. This capability is becoming increasingly important for edge AI services such as autonomous driving, mobile platforms, and IoT environments \cite{wang2025empowering, choi2024intelligent}. In such settings, pretrained LLMs often require domain-specific adaptation and continual updates from newly generated user data. Because this data frequently contains sensitive personal information, sending the raw data to a centralized cloud for fine-tuning is undesirable \cite{wang2025empowering, shin2026prediction}. Collaborative fine-tuning addresses this challenge by allowing training to proceed directly on edge devices while keeping the raw data local \cite{lin2024split}.

Existing studies on collaborative fine-tuning have mainly focused on distributing fine-tuning across edge devices \cite{ye2024asteroid, ouyang2024pluto, li2025dgpas, yoon2021edgepipe, choi2024harmonia}. Most target GPU-based edge platforms and rely on pipeline parallelism (PP), which improves device utilization and reduces iteration time by overlapping computation and communication. However, collaborative fine-tuning is also needed on CPU-based edge devices. In real deployments, many edge platforms, including Siemens’ IoT gateways \cite{siemens_iot2050}, smart-home hubs \cite{homeassistant_green}, and in-vehicle computers \cite{neousys_nuvo_2610vtc}, are primarily CPU-based. Even when GPUs or NPUs are available, they are often reserved for latency-critical inference rather than fine-tuning \cite{qualcomm_gpu,google_coral_npu}. As a result, many edge environments require collaborative fine-tuning on CPUs.

In this paper, we study collaborative fine-tuning on heterogeneous CPU-based edge devices, a setting that remains largely unexplored. We begin with a system-level analysis of PP-based techniques on CPU devices (\S\ref{sec:motivation}). Our analysis shows that the key advantage of PP, computation–communication overlap, does not hold in this setting. On CPU-based edge devices, the same processor handles both model computation and communication, causing resource contention. This contention prevents effective overlap and leads to an average 5.75$\times$ higher computation stall ratio than on GPU devices. These results indicate that directly applying GPU-oriented PP designs to CPU edge environments is inefficient.

We therefore explore an alternative strategy that reduces CPU contention. Tensor parallelism (TP) is a promising candidate because it does not rely on computation--communication overlap. Instead, TP separates computation and synchronization into sequential phases and improves training speed through tensor partitioning and collective communication. This structure makes TP a better fit for CPU-based edge devices, where reducing CPU contention is critical.

However, we find that directly applying existing TP techniques to heterogeneous CPU edge devices is still ineffective. Our experiment with Megatron-LM \cite{shoeybi2019megatron} shows that the fastest device is idle for up to 34\% of an iteration. In other words, more than one-third of the fast workers are waiting for synchronization rather than performing fine-tuning. This is not merely a local inefficiency: as TP proceeds synchronously, such waiting time directly translates into wasted compute capacity and longer iteration time for the whole system.
This inefficiency arises as most TP techniques are designed for homogeneous datacenter settings with GPUs of similar compute and memory capabilities, whereas edge devices exhibit significant computational heterogeneity.

To address these challenges, we propose \xronos, a collaborative fine-tuning system for LLMs on heterogeneous CPU-based edge devices. \xronos adopts TP as its backbone to avoid CPU contention. It also introduces heterogeneity-aware planning to minimize idle time caused by stragglers. Specifically, it profiles a small set of representative, non-repeated LLM layers and uses the results to predict per-device computation cost and memory usage. Based on these predictions, it searches for a partitioning strategy that reduces the idle time and thereby increases the training speed.

The key contributions of this study are:
\begin{itemize}[leftmargin=1.5em,itemsep=0pt,topsep=0pt]
    \item Present a system-level analysis of collaborative LLM fine-tuning on CPU-based edge devices and show why PP is ineffective in this setting.
    \item Design \xronos, a TP-based collaborative fine-tuning system that combines lightweight profiling with heterogeneity-aware tensor partitioning.
    \item Demonstrate that \xronos improves iteration time by up to 56\% and reduces idle time ratio by up to 5.9$\times$, across diverse device combinations and workloads.
\end{itemize}

\section{Background}

\subsection{Edge Collaborative Fine-tuning}

Edge collaborative fine-tuning updates a shared LLM across multiple trusted edge devices connected through a local network. One device acts as the coordinator, orchestrating training, profiling the participating devices when needed, and selecting a parallelization strategy for fine-tuning. Other devices act as workers. Throughout the paper, we use worker to denote a device that stores part or all of the model and participates in collaborative fine-tuning.

Fine-tuning proceeds over epochs and iterations. An epoch is one full pass over the fine-tuning dataset. Each epoch consists of multiple iterations, and each iteration processes one global batch. The global batch size is the total number of training examples processed by all workers in an iteration. Depending on the parallelization strategy, the global batch may be further subdivided; e.g., PP splits it into micro-batches.

In each iteration, each worker performs forward and backward computation on its assigned portion of the model, such as a pipeline stage in PP or a set of partitioned tensors in TP, and exchanges intermediate tensors (e.g., activations and gradients) with other workers over the network. The per-worker iteration time thus consists of local computation time and communication time. Execution is synchronous: the coordinator starts the iteration, and the system advances to the next iteration after all workers finish the current one. So, the end-to-end iteration time is determined by the slowest worker.

\begin{figure}[t]
  \centering
  \subfloat[Model partitioning.\label{fig:pp_model}]{
    \includegraphics[width=0.3 \columnwidth]{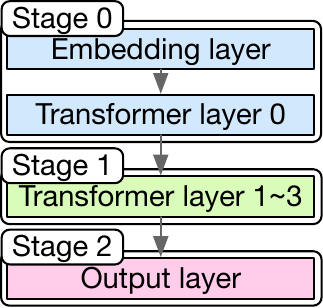}
  }
  \vspace{.4em}
  \subfloat[Execution workflow.\label{fig:pp_workflow}]{
    \includegraphics[width=0.6\columnwidth]{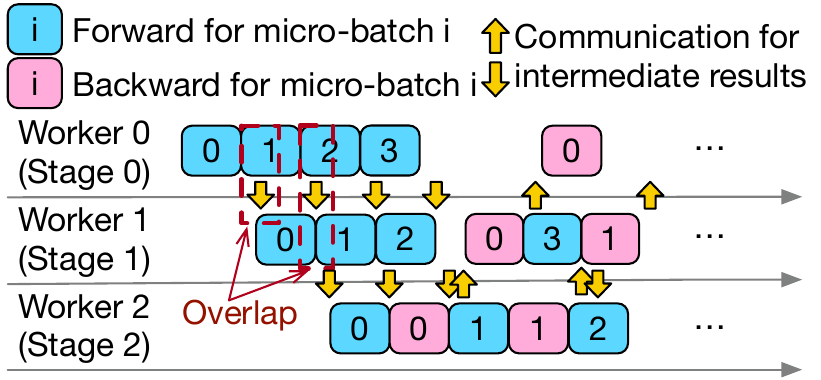}
  }\vspace{-.4em}
  \caption{PP example with three workers.}
  \label{fig:pp}\vspace{-1.8em}
\end{figure}

\subsection{Parallelization Strategy}\label{sec:back_strategy}

We focus on two representative strategies for collaborative fine-tuning on edge workers: PP and TP. Note that although other parallelization strategies are in principle possible, PP and TP capture the main design trade-offs relevant to our edge-device settings. To the best of our knowledge, prior edge collaborative fine-tuning systems have primarily built on PP, making it the most relevant baseline and TP the natural alternative to examine.

\subsubsection{PP}
PP divides the model into a sequence of stages, where each stage contains consecutive layers and is assigned to a single worker. Fig. \ref{fig:pp_model} shows a three-worker example. The model is partitioned from the embedding layer to the output layer into stage 0, stage 1, and stage 2, which are mapped to worker 0, worker 1, and worker 2, respectively.

During each iteration, PP splits the global batch into multiple micro-batches and pipelines them across the stages (Fig. \ref{fig:pp_workflow}). Each worker executes forward and backward passes only for its own stage. During the forward pass, a worker sends activation tensors to the next stage; during the backward pass, it sends gradient tensors to the previous stage. PP attempts to overlap this communication with computation across different micro-batches and stages. For example, while worker 1 in Fig. \ref{fig:pp_workflow} computes the forward pass of one micro-batch, it can simultaneously receive activations for another micro-batch from worker 0. By hiding part of the communication latency behind computation, PP aims to reduce the overall iteration time. Existing collaborative fine-tuning systems predominantly utilize PP \cite{ye2024asteroid, ouyang2024pluto, li2025dgpas, al2024optimizing}.

\subsubsection{TP}
Unlike PP, TP does not assign disjoint groups of layers to workers. Instead, TP partitions tensors within selected layers across workers; we refer to these as partitioned tensors. Throughout the rest of the paper, we call a layer \textit{partitioned} to be distributed across workers via tensor partitioning, and \textit{replicated} when the layer is fully maintained on every worker.

In TP, compute- and memory-intensive layers, such as embedding, self-attention, and linear layers, are partitioned, as their large tensors benefit from distributed storage and parallel computation. In contrast, layers and operations with small parameter sizes or limited parallelization benefits, such as layer normalization, dropout, and the final output layer, are typically replicated to avoid unnecessary synchronization overhead.

\begin{figure}[t]
  \centering
  \subfloat[Model partitioning.\label{fig:tp_model}]{
    \includegraphics[width=0.82\columnwidth]{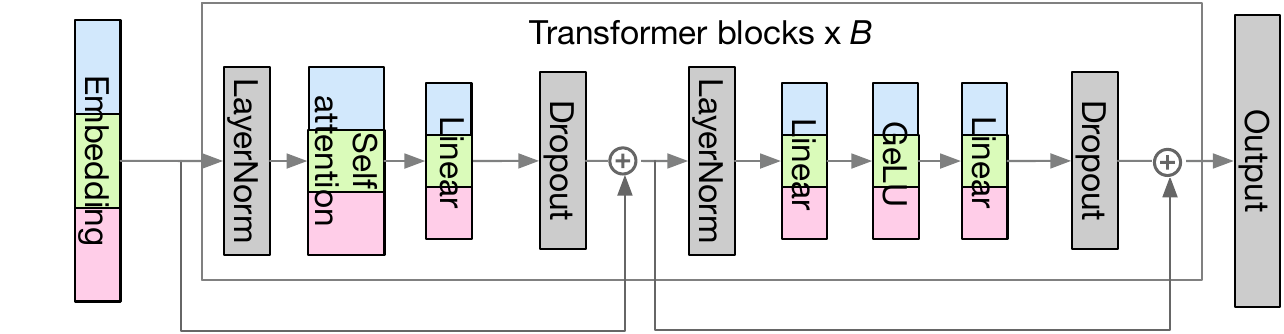}
  }
  \vspace{.1em}
  \subfloat[Execution workflow.\label{fig:tp_workflow}]{
    \includegraphics[width=0.93\columnwidth]{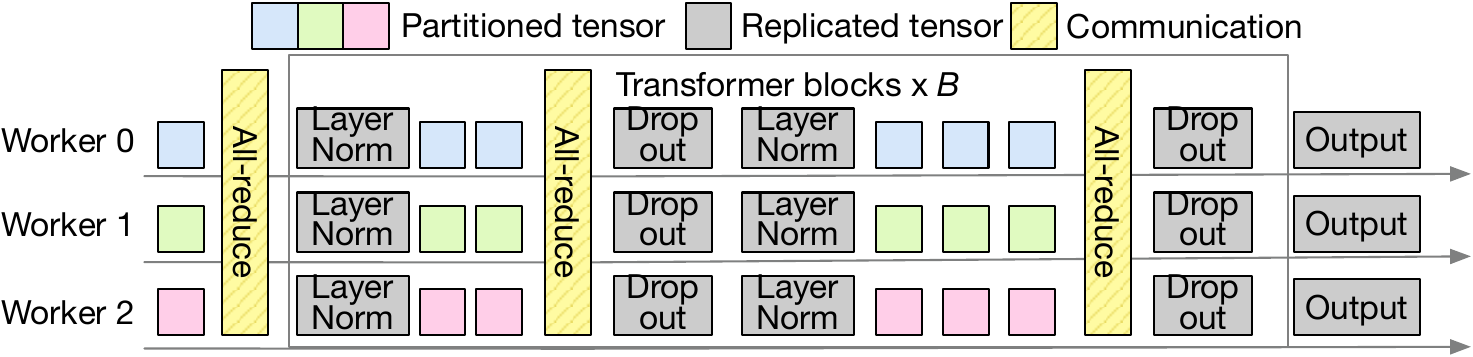}
  }\vspace{-.4em}
  \caption{TP example with three workers.}
  \label{fig:tp}\vspace{-1.9em}
\end{figure}

\begin{figure*}[t]
    \centering
    \captionsetup{justification=centering}

    \begin{minipage}[t]{0.24\textwidth}
        \centering
        \vspace{0pt}
        \includegraphics[height=2.4cm]{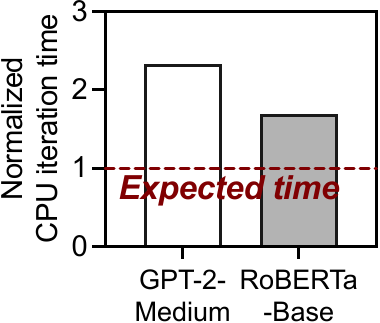}
        \vspace{-0.5em}
        \captionof{figure}{Normalized CPU iteration time of PP (\S\ref{sec:mot_pp_iter}).}
        \label{fig:moti}
    \end{minipage}%
    %
    \begin{minipage}[t]{0.21\textwidth}
        \centering
        \vspace{0pt}
        \captionof{table}{Computation stall ratio of PP (\S\ref{sec:mot_pp_stall}).}
        \label{tab:pp_profile}
        \setlength{\tabcolsep}{2pt}
        \scriptsize
        \begin{tabular}{lcc}
            \toprule
            & \textbf{\makecell{GPT-2- \\ Medium}} & \textbf{\makecell{RoBERTa \\ -Base}} \\
            \midrule
            \textbf{GPU worker} & 9.03\% & 12.63\% \\
            \textbf{CPU worker} & 65.52\% & 53.57\% \\
            \textbf{CPU/GPU} & 7.26$\times$ & 4.24$\times$ \\
            \bottomrule
        \end{tabular}
    \end{minipage}%
    %
    \begin{minipage}[t]{0.24\textwidth}
        \centering
        \vspace{0pt}
        \captionof{table}{CPU overheads between PP and TP (\S\ref{sec:exp_tp_cpu}).}
        \label{tab:tp_profile}
        \setlength{\tabcolsep}{2pt}
        \scriptsize
        \begin{tabular}{ccc}
            \toprule
            \textbf{Metric} & \textbf{Model} & \(\frac{\textbf{PP}-\textbf{TP}}{\textbf{PP}}\) \\
            \midrule
            \multirow{2}{*}{\makecell{Context\\switches}}
              & GPT-2-Medium & 95.02\% \\
              & RoBERTa-Base & 97.70\% \\
            \midrule
            \multirow{2}{*}{\makecell{Comp.\\stall}}
              & GPT-2-Medium & 19.98\% \\
              & RoBERTa-Base & 14.97\% \\
            \bottomrule
        \end{tabular}
    \end{minipage}%
    %
    \begin{minipage}[t]{0.31\textwidth}
        \centering
        \vspace{0pt}
        \begin{subfigure}[t]{0.5\linewidth}
            \centering
            \includegraphics[height=2.3cm]{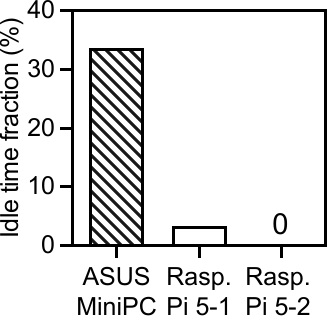}
            \caption{GPT-2-Medium}
            \label{fig:gpt2}
        \end{subfigure}\hfill
        \begin{subfigure}[t]{0.5\linewidth}
            \centering
            \includegraphics[height=2.3cm]{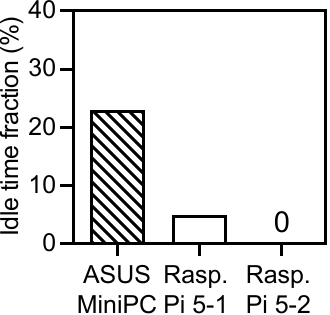}
            \caption{RoBERTa-Base}
            \label{fig:roberta}
        \end{subfigure}
        \vspace{-0.5em}
        \captionof{figure}{Idle time ratio of TP (\S\ref{sec:exp_tp_idle}).}
        \label{fig:idle_iteration}
    \end{minipage}
    \vspace{-1.7em}
\end{figure*}

Fig. \ref{fig:tp_model} shows a TP example with three workers. Colored blocks denote partitioned tensors distributed across workers, while gray blocks denote replicated layers. For each partitioned layer, every worker computes its assigned tensor partition in parallel. The workers then communicate (yellow boxes) to synchronize and aggregate partial results through collective communication (e.g., all-reduce) to produce the activations or gradients required for subsequent computation (Fig. \ref{fig:tp_workflow}). Replicated layers are executed independently on each worker between synchronization points. In contrast to PP that overlaps communication and computation across micro-batches, TP separates the two.

\section{System-level Analysis of Edge Collaborative Fine-tuning on CPU Devices}\label{sec:motivation}

This section addresses two questions. First, does PP retain its main benefit on CPU-based edge workers? Second, if PP is ineffective, can conventional TP serve as a direct replacement? To answer the questions, we first analyze PP on GPU and CPU workers and then examine TP on CPU workers with and without hardware heterogeneity. The results motivate the design of \xronos.

\subsection{Experiment Setup}\label{sec:exp_setup_motivation}

\subsubsection{Workload}
We fine-tune GPT-2-Medium \cite{radford2019language} and RoBERTa-Base \cite{liu2019roberta} in FP32 with a sequence length of 32 and a global batch size of 8. For PP, each global batch is split into 8 micro-batches to expose the intended pipeline overlap. We use Gloo \cite{gloo} for inter-worker communication and fine-tune the models on the CoLA task from the GLUE benchmark \cite{wang2018glue}. We use a single downstream task in this section, as the goal is to isolate system behavior rather than compare task accuracy. Unless otherwise stated, all results are averaged over the iterations within a single epoch.

\subsubsection{Comparison}
We compare representative PP and TP techniques. For PP, we use Asteroid~\cite{ye2024asteroid}, a state-of-the-art (SOTA) heterogeneity-aware pipeline parallelism technique. For TP, we use Megatron-LM~\cite{shoeybi2019megatron}, a widely used tensor parallelism technique for homogeneous GPU workers.

\subsubsection{Devices}
For the PP analysis in \S\ref{sec:exp_pp_ineffective}, we compare two scenarios: (1) three GPU workers of Jetson Orin Nano devices and (2) three CPU workers of Raspberry Pi 5 devices. By comparing GPU and CPU workers, we identify the problem of existing techniques on CPU workers.

For the TP analysis in \S\ref{sec:exp_tp_ineffective}, we compare two scenarios: (1) three homogeneous CPU workers (Raspberry Pi 5) and (2) three heterogeneous CPU workers (two Raspberry Pi 5 of four-core ARM Cortex-A76, 2.4 GHz and one ASUS MiniPC of four-core Intel N100, 3.4 GHz). We identify the problem of existing TP techniques for heterogeneous CPU workers.

\subsection{Analysis 1: PP Loses Its Main Advantage on CPU Workers}\label{sec:exp_pp_ineffective}

\subsubsection{Iteration time}\label{sec:mot_pp_iter}
We compare the average iteration time of PP on GPU and CPU workers. Iteration time is defined as the end-to-end latency of a single iteration, including both computation and communication, and is determined by the slowest worker under synchronous execution.

We first measure the execution time of a single transformer block from GPT-2-Medium and RoBERTa-Base on CPU and GPU workers without collaborative fine-tuning to quantify the compute speed difference between CPU and GPU. We observe that the transformer blocks take 3.35$\times$ and 2.51$\times$ longer on CPU workers than on GPU workers.

Next, we measure the iteration time during collaborative fine-tuning. To isolate the overhead beyond the inherent CPU–GPU speed difference, we normalize the CPU iteration time by (1) the single-block speed ratios (3.35$\times$ and 2.51$\times$) and (2) the GPU iteration time. A normalized value of 1 indicates that the CPU worker has the same iteration time as the GPU worker, after accounting for the difference in compute speed between the GPU and the CPU.

Fig. \ref{fig:moti} shows that the normalized CPU iteration time reaches 2.33 for GPT-2-Medium and 1.70 for RoBERTa-Base. This result indicates that PP introduces additional slowdown beyond the compute speed difference between CPU and GPU workers. We next investigate the cause of this slowdown.

\subsubsection{Computation–communication overlap}\label{sec:mot_pp_stall}
We next analyze why PP slows down disproportionately on CPU workers. We measure the computation stall ratio, defined as $(\text{backend stall cycles} / \text{CPU cycles}) \times 100$, measured through the Linux \texttt{perf\_event\_open} interface \cite{perf_event_open}. A higher value indicates that the worker spends a larger fraction of execution time stalled rather than making forward progress. 

Table \ref{tab:pp_profile} reports the computation stall ratio for PP on (1) GPU workers, (2) CPU workers, and (3) their comparison (CPU divided by GPU). CPU workers show, on average, 5.75$\times$ higher stall ratios than GPU workers---7.26$\times$ for GPT-2-Medium and 4.24$\times$ for RoBERTa-Base. So, PP in CPU workers poorly overlaps communication with computation.

The reason is architectural. On GPU workers, model computation runs on the GPU while communication is handled by the host CPU, so the two can proceed largely in parallel. On CPU workers, however, the same processor should execute both computation and communication. As a result, the overlap of PP turns into contention for the same CPU cores and runtime resources, greatly reducing the benefit of pipelining.



\begin{figure*}[t]
  \centering
  \includegraphics[width=.88\textwidth]{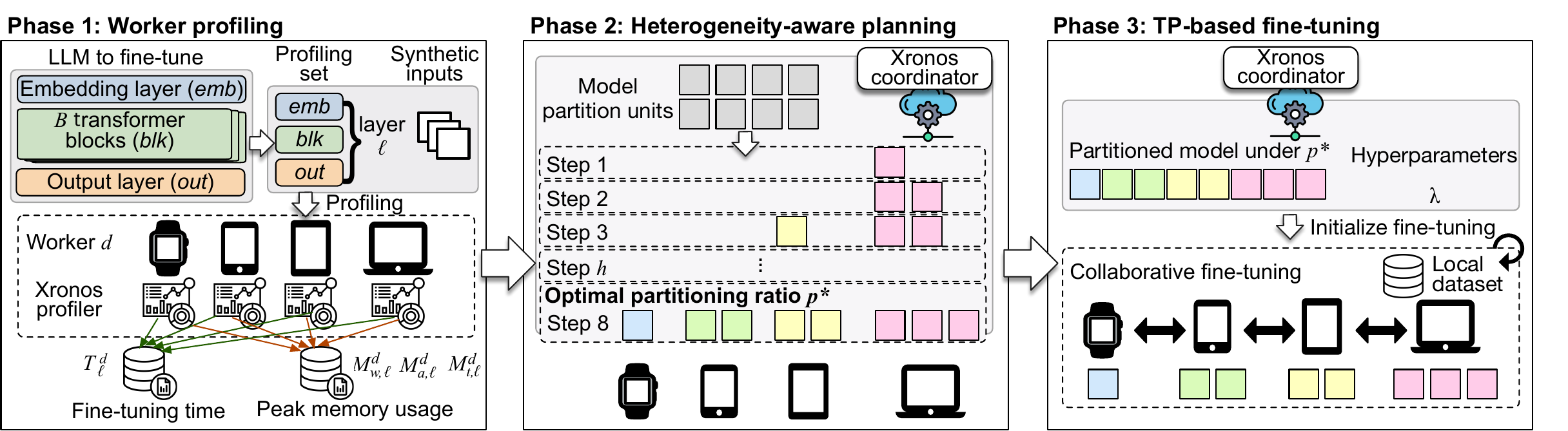}\vspace{-.2em}
  \caption{\xronos architecture and end-to-end workflow.}
  \label{fig:xronos_overview}\vspace{-1.8em}
\end{figure*}

\subsection{Analysis 2: Existing TP Remains Inefficient on Heterogeneous CPU Workers}\label{sec:exp_tp_ineffective}

We consider TP as an alternative, as it separates computation from communication (\S\ref{sec:back_strategy}) and is inherently better suited for CPU-based workers. We thus ask two questions: (1) does TP reduce CPU contention, and (2) if so, do existing TP techniques remain effective for CPU workers?

\subsubsection{CPU contention}\label{sec:exp_tp_cpu}
We compare TP and PP on three homogeneous CPU workers (Raspberry Pi 5). We measure two metrics: (1) the number of context switches and (2) the computation stall ratio, both obtained using \texttt{perf\_event\_open}. The number of context switches reflects OS scheduling overhead; a higher value indicates greater interference between computation and communication tasks.

Table \ref{tab:tp_profile} reports the relative reduction of TP compared to PP, computed as (PP$-$TP)/PP. TP reduces context switches by 95.02\% for GPT-2-Medium and 97.70\% for RoBERTa-Base. It also reduces the computation stall ratio by 19.98\% and 14.97\%, respectively. The reductions explain that, to mitigate CPU contentions, TP is a better backbone for collaborative fine-tuning on CPU workers than PP.

\subsubsection{Idle time ratio}\label{sec:exp_tp_idle}
However, lower CPU contention alone does not make existing TP techniques sufficient for CPU workers. We evaluate Megatron-LM, a representative TP technique, on three heterogeneous CPU workers (two Raspberry Pi 5 and one ASUS MiniPC). We measure the idle time ratio, defined as the fraction of an iteration during which a worker waits at synchronization points for other workers to catch up.

Fig. \ref{fig:idle_iteration} shows that the two Raspberry Pi 5 workers have idle time ratios below 5\% for both models, indicating that they are almost fully utilized throughout the iteration. In contrast, the faster ASUS MiniPC worker remains idle for 34\% of the iteration for GPT-2-Medium and 23\% for RoBERTa-Base. For GPT-2-Medium, more than one-third of the fastest worker’s iteration time is spent waiting rather than performing useful computation. Because TP proceeds synchronously, this waiting time not only affects local workers but also directly increases the end-to-end iteration time.

The high idle ratios are due to the design assumptions of existing TP techniques. Most TP techniques are developed for relatively homogeneous GPU clusters and thus rely on uniform assignment of partitioned tensors \cite{shoeybi2019megatron, xu2023efficient, shi2025tapas}. With heterogeneous CPU workers, however, such uniform assignment creates stragglers: slower workers receive similar amounts of work despite their differences in compute capability, while faster workers repeatedly wait at synchronization points.

In summary, our analysis in this section suggests that an efficient collaborative fine-tuning system for CPU workers should satisfy two requirements: (1) it should avoid CPU contention inherent in PP, and (2) it should account for worker heterogeneity when determining the amount of computation assigned to each worker to reduce idle time.

\section{\xronos Design}

Fig. \ref{fig:xronos_overview} shows the \xronos architecture and its end-to-end workflow, which consists of three phases: (1) worker profiling, (2) heterogeneity-aware planning, and (3) TP-based fine-tuning.
In the profiling phase, each worker runs the \xronos profiler to measure its layer-wise fine-tuning time and peak memory usage, and then sends the results to the coordinator. In the planning phase, the coordinator collects these profiling results and derives TP strategy that matches each worker's compute and memory capacity. In the fine-tuning phase, the coordinator materializes the selected tensor layout and launches collaborative fine-tuning across the workers.

\subsection{Worker Profiling} \label{sec:profiling}

To capture worker heterogeneity with low overhead, \xronos profiles a small representative subset of the target LLM rather than the entire model. We decompose the model into three components: an embedding layer (\textit{emb}), $B$ transformer blocks (\textit{blk}), and a final output layer (\textit{out}). While \textit{emb} and \textit{out} are structurally similar across many LLMs, the internal composition of a transformer block depends on the model family. For example, a GPT-2 block contains ten layers---one self-attention layer, three linear layers, two layer-normalization layers, three dropout layers, and one GELU activation---whereas a LLaMA block contains eight layers---one self-attention layer, four linear layers, two layer-normalization layers, and one SiLU activation\cite{lu2025demystifying}.

Given an LLM to fine-tune, the \xronos profiler builds a profiling set consisting of one \textit{emb}, one representative \textit{blk}, and one \textit{out}. Let $k$ denote the number of layers inside the representative transformer block. The profiling set therefore contains $k+2$ layers. Instead of profiling the full model, the profiler runs a few iterations (e.g., 10) of fine-tuning on the profiling set, which we empirically find sufficient to accurately estimate device capabilities across different LLMs. We use synthetic inputs \cite{zhong2024distserve} because the goal of this phase is to characterize performance rather than task accuracy. 

During profiling, worker $d$ records two quantities for each layer $\ell$: (1) layer-wise fine-tuning time and (2) peak memory usage. First, let $T_{\ell}^{d}$ denote the time required for worker $d$ to run layer $\ell$ once. Because the profiling set contains $k+2$ layers, profiling yields $k+2$ values of $T_{\ell}^{d}$ per worker.

Second, let $M_{\ell}^{d}$ denote the peak memory usage of layer $\ell$ on worker $d$. We decompose this quantity into three separate parts: memory for model weights, activations, and temporary kernel buffers, denoted by $M_{w,\ell}^{d}$, $M_{a,\ell}^{d}$, and $M_{t,\ell}^{d}$, respectively. Thus, for each of the $k+2$ layers, the profiler records three memory metrics---$M_{w,\ell}^{d}$, $M_{a,\ell}^{d}$, and $M_{t,\ell}^{d}$---resulting in $3(k+2)$ distinct metrics per worker.

Note that we exclude communication time from our profiling metrics because it remains constant across different partitioning ratios. In TP, workers synchronously exchange activation and gradient tensors after computing their local partitions, so the communication time is bounded by the slowest worker. Let $B_d$ denote the network bandwidth of worker $d$. The communication time $T_{\mathrm{comm}}$ becomes $V_{\mathrm{comm}} / {\min\limits_{d \in \mathcal{D}} B_d}$. Here, $V_{\mathrm{comm}}$ denotes the communication volume, which is determined only by the batch size and the model’s original hidden dimension~\cite{shoeybi2019megatron}. So, it remains consistent across workers and partitioning ratios. The communication time is thus determined by $\min\limits_{d \in \mathcal{D}} B_d$, i.e., the minimum worker bandwidth, which is also independent of the partitioning ratio and consistent for a given set of edge workers \cite{devraj2025efficient}. Therefore, $T_{\mathrm{comm}}$ remains constant across partitioning ratios and is not profiled separately.

\subsection{Heterogeneity-aware Planning} \label{sec:planning}

Using the profiles, the coordinator determines how many partitioned tensors to place on each worker. As described in \S\ref{sec:back_strategy}, each worker stores (1) replicated tensors for layers that are not partitioned and (2) a worker-specific fraction of partitioned tensors. The coordinator chooses these fractions to reduce straggler-induced waiting while respecting per-worker memory limits.

\subsubsection{Partition ratio}
Let $p_d$ denote the fraction of all partitioned tensors assigned to worker $d$.
In TP, tensors cannot be split arbitrarily; instead, they are assigned in indivisible chunks that must reside on a single worker. We call such a chunk a partition unit. For example, in self-attention, all tensors associated with one attention head form a partition unit, since they jointly produce one output.
Accordingly, TP assigns partition units to each worker in integer numbers \cite{shoeybi2019megatron}. In LLMs, each layer contains the same number of partition units, $H$, which defines the granularity of workload distribution.

If worker $d$ receives $h$ partition units, its partition ratio is $p_d = h/H$. The feasible set of partition ratios is then:
\vspace{-.2em}\begin{equation}
\label{eq:split_ratio}
p_d \in \mathbb{P}=\left\{{h}/{H}\;\middle|\; h\in\{0,1,\dots,H\}\right\}.\vspace{-.2em}
\end{equation}

\subsubsection{Problem formulation}
Suppose that $N$ workers are available, where worker $d \in \{1,\dots,N\}$ has memory capacity $M_{cap}^{d}$. Our goal is to choose partition ratios $\mathbf{p}^* = (p_1^*, \dots, p_N^*)$ that minimize the bottleneck in worker time. Under the TP setting we target, synchronization cost is identical across workers for a given model and batch configuration, so the worker with the longest local fine-tuning time determines the end-to-end iteration time. We therefore formulate the problem:
\vspace{-.4em}\begin{align}
\mathbf{p}^* = \mathop{\arg\min}\limits_{\{p_1, \dots, p_N\}} & \left[ \max_{1 \le d \le N} \left\{ T^{d}(p_d) \right\} \right] \label{eq:opt_problem} \\
\textrm{s.t.} \quad 
& \textstyle\sum_{d=1}^{N} p_d = 1 \label{eq:constraint_sum} \\
& M_{total}^{d}(p_d) \le M_{cap}^{d}, \quad \forall d \in \{1, \dots, N\} \label{eq:constraint_mem} \\
& p_d \in \mathbb{P}, \quad \forall d \in \{1, \dots, N\}. \label{eq:constraint_granularity}\vspace{-.4em}
\end{align}

Eq. \eqref{eq:opt_problem} minimizes the maximum fine-tuning time across workers, where $T^{d}(p_d)$ denotes the fine-tuning time of worker $d$ under partition ratio $p_d$. Eq. \eqref{eq:constraint_sum} ensures that all tensors to be partitioned are fully distributed across workers. Eq. \eqref{eq:constraint_mem} enforces the memory constraint of each worker, where $M_{total}^{d}(p_d)$ denotes the total memory usage of worker $d$. Finally, Eq. \eqref{eq:constraint_granularity} restricts each $p_d$ to the discrete candidate set $\mathbb{P}$ defined in Eq. \eqref{eq:split_ratio}.

\subsubsection{Partition-aware cost estimation}
To solve Eq. \eqref{eq:opt_problem}--Eq. \eqref{eq:constraint_granularity}, \xronos estimates two quantities for each worker: its fine-tuning time $T^{d}(p_d)$ and its peak memory usage $M_{total}^{d}(p_d)$. Both are derived from the profiling results obtained in \S\ref{sec:profiling}.

\textbf{$\boldsymbol{T^d(p_d)}$ estimation.}
We estimate the worker-side fine-tuning time as
\begin{equation}
\label{eq:total_comp_time}
T^{d}(p_d) = p_d \cdot T_{emb}^{d} + T_{out}^{d} + B \cdot T_{blk}^{d}(p_d).
\end{equation}

The first term scales the embedding-layer cost by $p_d$ because worker $d$ stores only a fraction $p_d$ of the embedding tensors.\footnote{As the embedding computation is proportional to the number of tensor elements, the execution time scales linearly with the tensor size.} The second term is independent of $p_d$ because the output layer is replicated on every worker. The last term captures the cost of the $B$ transformer blocks, where
\vspace{-.2em}\begin{equation}
\label{eq:per_layer_comp_time}
T_{blk}^{d}(p_d)
= \textstyle\sum_{\ell \in blk_{part}} p_d \cdot T_{\ell}^{d}
+ \sum_{\ell \in blk_{repl}} T_{\ell}^{d}.\vspace{-.2em}
\end{equation}

Here, the first sum corresponds to partitioned layers within one transformer block and scales with $p_d$, whereas the second sum corresponds to replicated layers and is independent of $p_d$. All $T_{l}^{d}$ values are known from the worker profiling results.

\textbf{$\boldsymbol{M_{total}^{d}(p_d)}$ estimation.}
The total peak memory usage is the sum of (1) weights $M_{w}^{d}(p_d)$, (2) activations $M_{a}^{d}(p_d)$, and (3) temporary buffer memory $M_{t}^{d}(p_d)$ under $p_d$:
\vspace{-.2em}\begin{equation}
M_{total}^{d}(p_d) 
= M_{w}^{d}(p_d) 
+ M_{a}^{d}(p_d) 
+ M_{t}^{d}(p_d).
\label{eqy}\vspace{-.2em}
\end{equation}

First, $M_{w}^{d}(p_d)$ is derived as:
\begin{equation}
\label{eq:mem_static}
\begin{aligned}
M_{w}^{d}(p_d)
&= p_d \cdot M_{w,emb}^{d}
+ M_{w,out}^{d} \\
&\quad + B \left(
p_d {\textstyle\sum_{\ell\in {blk}_{part}}} M_{w,\ell}^{d}
+ \textstyle\sum_{\ell\in{blk}_{repl}} M_{w,\ell}^{d}
\right).
\end{aligned}
\end{equation}

Weights remain resident in memory throughout the entire fine-tuning. So, the total static memory usage is obtained by summing the embedding, output, and $B$ transformer block tensors, where the embedding tensors and partitioned tensors in each transformer block scale with $p_d$, while replicated tensors are independent of $p_d$.
Similar to the weights, activations also remain in memory throughout the fine-tuning process. Thus, $M_a^d(p_d)$ follows the same structure as Eq.\eqref{eq:mem_static}, replacing each weight term $M_w$ with the corresponding activation term $M_a$.




Lastly, $M_{t}^{d}$ refers to the peak memory usage from temporary buffers used to store intermediate results during computation (e.g., matrix multiplications or gradient computations). 
Unlike weights or activations above, the buffers are short-lived and exist only during the execution of a specific tensor computation. 
They are released immediately upon completion of computation and reused by subsequent operations. 
So, temporary buffers do not accumulate across layers, and the peak temporary memory is determined by the largest buffer during one fine-tuning iteration, which is calculated as follows:
\begin{equation}
\label{eq:mem_peak}
\begin{aligned}
M_{t}^{d}(p_d)
= \max \Big\{ \,
& p_d \cdot M_{t,emb}^{d},\  M_{t,out}^{d}, \\
& \max_{\ell \in {blk}_{part}} \left( p_d \cdot M_{t,\ell}^{d} \right),\ 
\max_{\ell \in {blk}_{repl}} \left( M_{t,\ell}^{d} \right)
\Big\}.
\end{aligned}
\end{equation}

\subsubsection{TP strategy search}\label{sec:search}
We search for the TP strategy to determine the final partition ratios $\mathbf{p}^*$, by solving Eq. \eqref{eq:opt_problem}--\eqref{eq:constraint_granularity}. Finding the global optimum requires exploring all possible allocations of $H$ partition units across $N$ workers. However, the resulting search space of size $\binom{H+N-1}{N-1}$ leads to exponential time complexity, which quickly becomes prohibitive even for moderate $H$ and $N$. To make the search practical, we use a greedy partition-unit assignment strategy that allocates one partition unit at a time, reducing the number of candidate evaluations to $O(HN)$.

The search starts from an empty partition assignment, i.e., $\mathbf{p}^*=\mathbf{0}$, and considers all workers as candidates for receiving partition units. At each step, \xronos tentatively assigns one additional partition unit, $\Delta p=1/H$, to each candidate worker and verifies whether the resulting partition ratio satisfies the worker's memory constraint. Workers that cannot accommodate the additional unit are excluded from further consideration.

The partition unit is then assigned to the feasible worker that yields the smallest estimated fine-tuning time after receiving the unit. Because TP execution is synchronous, the overall iteration time is determined by the slowest worker. By always assigning to the fastest worker, the search balances the load across workers and directly reduces the bottleneck. 

The process continues until all $H$ units are assigned. If all partition units are successfully assigned, the resulting ratios $\mathbf{p}^*$ are used as the TP strategy for fine-tuning. If no feasible worker remains before all units are allocated, the search returns \texttt{null}, indicating that the available workers do not provide enough aggregate memory to host the target model.

\subsection{TP-based Fine-tuning}

After searching $\mathbf{p}^*$, the \xronos coordinator materializes the TP layout by placing replicated and partitioned tensors on workers according to $\mathbf{p}^*$. It then initializes collaborative fine-tuning, including the hyperparameters (e.g., global batch size, sequence length, learning rate, and optimizer) and dataset. Once initialization finishes, the coordinator launches synchronized TP-based fine-tuning across the workers.

\begin{table}[t!]
    \centering
    \begin{minipage}{.5\textwidth}
        \centering
        \caption{Model specifications.}
        \label{tab:model_specs}\vspace{-.5em}
        \renewcommand{\arraystretch}{.8}
        \begin{tabular}{lccc}
            \toprule
            \textbf{Model} & \textbf{Arch.} & \textbf{Params} & \textbf{\# Transformer blocks} \\
            \midrule
            RoBERTa-Base & Encoder & 125M & 12 \\
            GPT-2-Medium & Decoder & 345M & 24 \\
            MobileLLaMA-1.4B & Decoder & 1.4B & 24 \\
            \bottomrule
        \end{tabular}
    \end{minipage}\vspace{1em}
    \begin{minipage}{.5\textwidth}
        \centering
        \caption{Device specifications.}
        \label{tab:hardware_specs}\vspace{-.5em}
        \renewcommand{\arraystretch}{.8}
        \begin{tabular}{lcc}
            \toprule
            \textbf{Device} & \textbf{CPU processor} & \textbf{Memory} \\
            \midrule
            Raspberry Pi 5 & ARM A76 (2.4GHz) & 8 GB \\
            Orange Pi 5+ & ARM A76/A55 (2.4/1.8GHz) & 16 GB \\
            LattePanda Mu & Intel N100 (3.4GHz)  & 8 GB \\
            ASUS MiniPC & Intel N100 (3.4GHz) & 16 GB \\
            \bottomrule
        \end{tabular}
    \end{minipage}\vspace{-1.8em}
\end{table}

\begin{figure*}[h!]
    \centering
    \captionsetup[subfigure]{justification=centering,singlelinecheck=false}

    \includegraphics[width=0.35\textwidth]{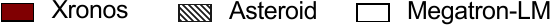}
    \vspace{-.1em}

    \begin{subfigure}[t]{0.41\textwidth}
        \centering
        \includegraphics[height=2.2cm]{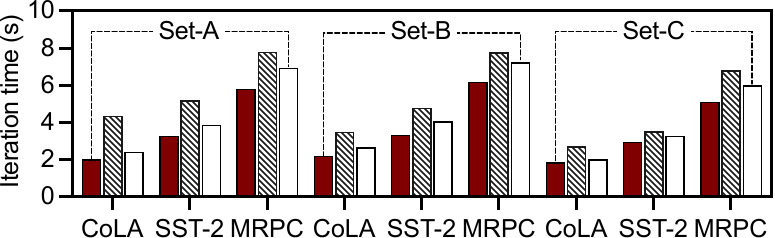}
        \caption{RoBERTa-Base}
        \label{fig:roberta-iter}
    \end{subfigure}\hfil
    \begin{subfigure}[t]{0.385\textwidth}
        \centering
        \includegraphics[height=2.2cm]{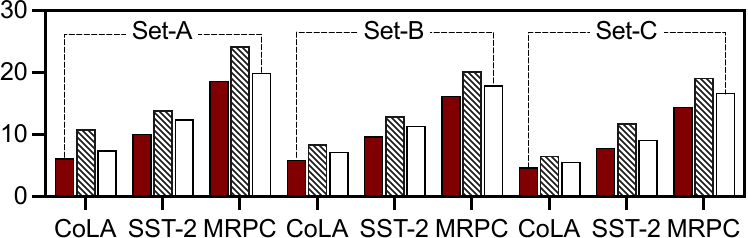}
        \caption{GPT-2-Medium}
        \label{fig:gpt-iter}
    \end{subfigure}\hfil
    \begin{subfigure}[t]{0.19\textwidth}
        \centering
        \includegraphics[height=2.2cm]{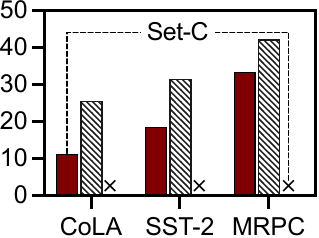}
        \caption{MobileLLaMA-1.4B}
        \label{fig:llama-iter}
    \end{subfigure}
    \vspace{-.3em}
    \caption{Iteration time comparison. $\times$ marks: out-of-memory error (\S\ref{sec:iteration-time}).}
    \vspace{-1.5em}
    \label{fig:iteration_time}
\end{figure*}

\section{Evaluation}\label{sec:eval}
We implement \xronos in PyTorch with $\sim$2K lines of code. For inter-worker communication, we use Gloo backend \cite{gloo} and PyTorch all-reduce operations \cite{torchdistributed}. 

\subsection{Experiment Setup}

\subsubsection{Comparison} We compare \xronos with two representative SOTA techniques: Asteroid \cite{ye2024asteroid} and Megatron-LM \cite{shoeybi2019megatron}.
Asteroid is a PP-based collaborative fine-tuning technique for edge environments, which partitions the model according to device compute capability and memory capacity. 
Megatron-LM is a widely used TP technique designed for high-end GPU clusters. 
It distributes computation uniformly across workers without accounting for device heterogeneity. 
We include Megatron-LM to evaluate the impact of heterogeneity-aware planning in \xronos. While other techniques exist \cite{ouyang2024pluto,li2025dgpas, shi2025tapas}, many are either not publicly available or follow designs similar to the two; thus, we use the two as baselines.

\subsubsection{Workloads}
We use three LLMs: RoBERTa-Base, GPT-2-Medium, and MobileLLaMA-1.4B \cite{chu2023mobilevlm}. Their architectures, parameter sizes, and number of transformer blocks are summarized in Table \ref{tab:model_specs}. We select the three because: 1) edge fine-tuning typically targets models under 2B parameters \cite{liu2024mobilellm}, and 2) they cover a wide range of model scales, from 125M (RoBERTa-Base) to 1.4B (MobileLLaMA-1.4B).
 
We fine-tune the models on three GLUE benchmark tasks \cite{wang2018glue}: CoLA (linguistic acceptability), SST-2 (sentiment analysis), and MRPC (paraphrase detection). We set the sequence lengths to 32, 64, and 128 for CoLA, SST-2, and MRPC, respectively. The minimum sequence length required to preserve model accuracy varies across tasks 
\cite{goyal2020power}; our choices fall within the ranges while covering diverse lengths. 
All experiments use FP32 precision with a global batch size of 8. For PP, each global batch is divided into 8 micro-batches. We use AdamW with a learning rate of $2\times10^{-5}$.
\subsubsection{Devices}
We evaluate the following three scenarios:
\begin{itemize}[leftmargin=1em,itemsep=0pt,topsep=0pt]
    \item Set-A (three devices): one ASUS MiniPC and two Raspberry Pi 5 devices.
    \item Set-B (four devices): one LattePanda Mu, one Orange Pi 5+, and two Raspberry Pi 5 devices.
    \item Set-C (five devices): one ASUS MiniPC, one LattePanda Mu, one Raspberry Pi 5, and two Orange Pi 5+ devices.
\end{itemize}
Table \ref{tab:hardware_specs} summarizes the specifications of each device type. 
The devices are heterogeneous in architecture, i.e., ARM (Raspberry Pi 5, Orange Pi 5+) and x86 (LattePanda Mu, ASUS MiniPC), with CPU frequencies ranging from 2.4 to 3.4 GHz and memory capacities from 8 to 16 GB. The differences result in heterogeneity in both compute capability and memory capacity across scenarios. All devices are connected via 1 Gbps Ethernet, similar to other studies \cite{ouyang2024pluto,bartolomeo2023oakestra}.
\subsubsection{Metrics}
We measure and report the following items:
\begin{itemize}[leftmargin=1em,itemsep=0pt,topsep=0pt]
\item Main results: we report the following metrics.
    \begin{itemize}
        \item Iteration time: average time per iteration.
        \item Computation stall ratio: amount of backend stall cycles divided by entire CPU cycles (\S\ref{sec:mot_pp_stall}).
        \item Number of context switches: total number of context switches (\S\ref{sec:exp_tp_cpu}) compared to PP (Asteroid).
        \item Idle time ratio: fraction of an iteration during which a device remains idle (\S\ref{sec:exp_tp_idle}).
    \end{itemize}
    \item Micro-benchmarks: we report the following metrics.
    \begin{itemize}
        \item Estimation error of fine-tuning time: the percentage error calculated by MAPE between the estimated value (\S\ref{sec:planning}) and the measured ground-truth.
    \item Accuracy--time analysis: fine-tuning accuracy over elapsed time. We compare time-to-accuracy, the time required to reach a target accuracy (e.g., 90\%).        
        \end{itemize}
\end{itemize}

Unless otherwise noted, all metrics are averaged over the iterations within a single epoch. Among the three models, we report all metrics for RoBERTa-Base and GPT-2-Medium across all scenarios. For MobileLLaMA-1.4B, we report results only for Set-C, as the others cannot load the model due to limited memory. The remaining setup follows \S\ref{sec:exp_setup_motivation}.

\subsection{Main results}
\label{sec:eval-main}
\subsubsection{Iteration time}\label{sec:iteration-time}
Figs. \ref{fig:roberta-iter}, \ref{fig:gpt-iter}, and \ref{fig:llama-iter} present the iteration time (y-axis) on three benchmark tasks (x-axis) for RoBERTa-Base, GPT-2-Medium, and MobileLLaMA-1.4B, respectively. The bars are grouped by scenarios (Set-A, Set-B, and Set-C).

Across all experiment cases, \xronos achieves the best (lowest) iteration time. For RoBERTa-Base in Fig. \ref{fig:roberta-iter}, \xronos reduces iteration time by 31\% (up to 53\% on Set-A of CoLA task) and 14\% (up to 18\% on Set-B of SST-2) on average compared to Asteroid and Megatron-LM, respectively. For GPT-2-Medium, \xronos reduces the iteration time by 28\% (up to 43\% on Set-A of CoLA) and 14\% (up to 18\% on Set-A of SST-2) on average. Also, for MobileLLaMA-1.4B, \xronos improves iteration time by 39\% (up to 56\% on CoLA) on average compared to Asteroid. Note that Megatron-LM fails due to an out-of-memory error because it does not account for device heterogeneity, including memory capacity. The results show that \xronos is more effective in terms of fine-tuning speed than existing techniques.

\begin{figure}[t!]
    \centering
    \captionsetup[subfigure]{justification=centering,singlelinecheck=false}

    \includegraphics[width=0.18\textwidth]{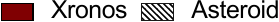}   
    \vspace{0.1em}

    \begin{subfigure}[t]{0.38\linewidth}
        \centering
        \includegraphics[height=2cm]{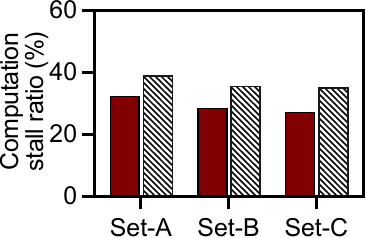}
        \caption{RoBERTa-Base}
        \label{fig:roberta-stall}
    \end{subfigure}\hfill
    \begin{subfigure}[t]{0.33\linewidth}
        \centering
        \includegraphics[height=2cm]{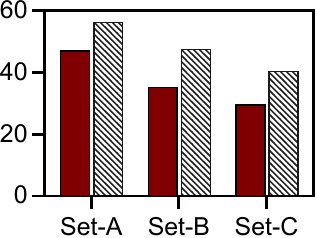}
        \caption{GPT-2-Medium}
        \label{fig:gpt-stall}
    \end{subfigure}
    \begin{subfigure}[t]{0.24\linewidth}
        \centering
        \includegraphics[height=2cm]{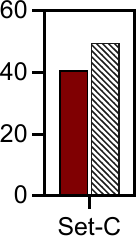}
        \caption{Mobile\\LLaMA-1.4B}
        \label{fig:llama-stall}
    \end{subfigure}
    \vspace{-.5em}
    \caption{Computation stall ratio comparison (\S\ref{sec:eval-stall-ratio}).}
    \label{fig:idle_time}\vspace{-.5em}
\end{figure}

\begin{table}[t]
\centering
\caption{Context switches comparison. All values are averaged across all devices and tasks. (\S\ref{sec:eval-context-switch})}
\label{tab:context_switch}
\begin{tabular}{llccc}
\toprule
\textbf{Scenario} & \textbf{Method} & \textbf{\shortstack[c]{RoBERTa\\-Base}} & \textbf{\shortstack[c]{GPT-2\\-Medium}} & \textbf{\shortstack[c]{MobileLLaMA\\-1.4B}} \\
\midrule
\multirow{2}{*}{Set-A}
& Xronos   & \textbf{8,190}  & \textbf{20,595} & -- \\
& Asteroid & 1,225,716       & 1,029,182       & -- \\
\midrule
\multirow{2}{*}{Set-B}
& Xronos   & \textbf{17,498} & \textbf{41,145} & -- \\
& Asteroid & 1,108,873       & 1,064,570       & -- \\
\midrule
\multirow{2}{*}{Set-C}
& Xronos   & \textbf{15,023} & \textbf{44,540} & \textbf{38,942} \\
& Asteroid & 1,166,493       & 1,133,191       & 1,240,182 \\
\bottomrule
\end{tabular}\vspace{-1.7em}
\end{table}

\begin{figure*}[t]
    \centering
    \captionsetup{justification=centering}
    \begin{minipage}[t]{0.48\textwidth}
        \vspace{0pt}
        \centering
        \captionsetup[subfigure]{justification=centering,singlelinecheck=false}
        \includegraphics[width=0.4\linewidth]{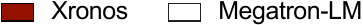}   
        \vspace{-.1em}
    
        \begin{subfigure}[t]{0.47\linewidth}
            \centering
            \includegraphics[height=2.5cm]{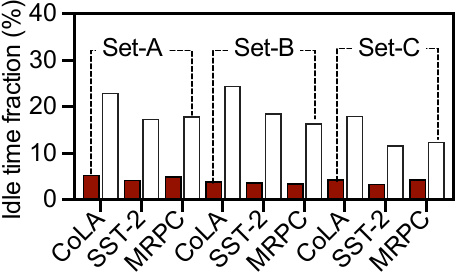}
            \caption{RoBERTa-Base}
            \label{fig:roberta-idle}
        \end{subfigure}
        \begin{subfigure}[t]{0.47\linewidth}
            \centering
            \includegraphics[height=2.5cm]{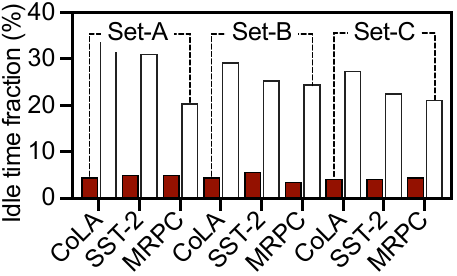}
            \caption{GPT-2-Medium}
            \label{fig:gpt-idle}
        \end{subfigure}
        \vspace{-.5em}
        \caption{Idle time ratio comparison (\S\ref{sec:eval-idle}).}
        \label{fig:idle_time}\vspace{-1em}
    \end{minipage}\hfill
    \begin{minipage}[t]{0.24\textwidth}
        \vspace{0pt}
        \centering
        \includegraphics[height=2.6cm]{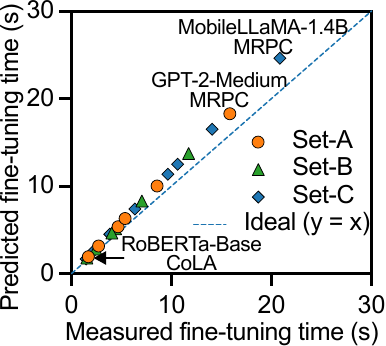}\captionsetup{justification=centering}
        \caption{Estimation accuracy (\S\ref{sec:eval-prediction}).}
        \label{fig:prediction}
    \end{minipage}\hfill
    \begin{minipage}[t]{0.24\textwidth}
        \vspace{0pt}
        \centering
        \includegraphics[height=2.6cm]{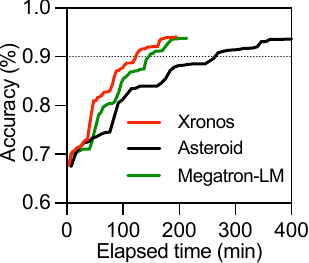}\captionsetup{justification=centering}
        \caption{Accuracy-time analysis (\S\ref{sec:eval-acc}).}
        \label{fig:eval-acc}
    \end{minipage}\vspace{-1.5em}
\end{figure*}



\subsubsection{Computation stall ratio} \label{sec:eval-stall-ratio}
Figs. \ref{fig:roberta-stall}, \ref{fig:gpt-stall}, and \ref{fig:llama-stall} show the computation stall ratio (y-axis) of \xronos and Asteroid for RoBERTa-Base, GPT-2-Medium, and MobileLLaMA-1.4B, respectively.
For each scenario (Set-A, Set-B, and Set-C), we measure the stall ratio for every device on each task (CoLA, SST-2, and MRPC) and report the average across all devices and tasks. 
Note that MobileLLaMA-1.4B reports values only for Set-C, as it can run only in this scenario, where the memory of devices is sufficient to load the model.
\xronos achieves lower computation stall ratios than Asteroid---on average, 20\%, 23\%, and 18\% for RoBERTa-Base, GPT-2-Medium, and MobileLLaMA-1.4B, respectively.

\subsubsection{Context switches} \label{sec:eval-context-switch}
Table~\ref{tab:context_switch} shows the number of context switches. 
Each value is measured in the same way as the computation stall ratio above. 
\xronos significantly reduces the number of context switches compared to Asteroid. 
Specifically, the reduction is 98.82\%, 96.73\%, and 96.86\% on average for RoBERTa-Base, GPT-2-Medium, and MobileLLaMA-1.4B.
The improved computation stall ratio and reduced context switches show that \xronos effectively alleviates CPU contention in PP by utilizing TP on CPU devices.

\subsubsection{Idle time ratio}\label{sec:eval-idle}

Figs. \ref{fig:roberta-idle} and \ref{fig:gpt-idle} show the idle time ratio of \xronos and Megatron-LM for RoBERTa-Base and GPT-2-Medium, respectively, on three tasks (x-axis). We exclude MobileLLaMA-1.4B because Megatron-LM fails to fine-tune it due to out-of-memory errors. 
\xronos maintains the idle time ratio below 6\% across both models and all configurations. In contrast, Megatron-LM shows $\sim$24\% and $\sim$34\% idle time ratios for RoBERTa-Base and GPT-2-Medium (with averages of 18\% and 26\%), respectively. This corresponds to a 4.6$\times$ and 5.9$\times$ reduction in idle time with \xronos.
Overall, \xronos consistently shows higher utilization on heterogeneous CPU workers than existing techniques in our setting.

\subsection{Micro-benchmarks}

\subsubsection{Time estimation error}\label{sec:eval-prediction}

Fig. \ref{fig:prediction} presents the estimated fine-tuning time (y-axis) against the measured ground-truth fine-tuning time (x-axis) across scenarios. Each point represents one LLM-task pair in each scenario and is measured using the optimal partition ratio $\mathbf{p}^*$ chosen by \xronos. As other partition ratios exhibit similar trends, we report results with $\mathbf{p}^*$ as representative. The diagonal line denotes the ideal estimation case ($y=x$), where the estimated fine-tuning time exactly matches the measured time. Thus, points closer to this line indicate higher estimation accuracy.

Most points lie close to the diagonal line---the average estimation error is 15.32\%, ranging from 13.28\% (RoBERTa-Base on CoLA in Set-A) to 16.65\% (MobileLLaMA-1.4B on MRPC in Set-C). The results indicate that, although the estimates are not perfectly accurate, using profiling results from the profiling set yields sufficiently reliable estimates to improve fine-tuning time, as explained in \S\ref{sec:planning}.

\subsubsection{Accuracy-time analysis}\label{sec:eval-acc}
Fig.~\ref{fig:eval-acc} presents the accuracy achieved during five epochs of collaborative fine-tuning. We report RoBERTa-Base on the CoLA dataset under the Set-A scenario as a representative case; other settings show similar trends. \xronos reaches the target accuracy in the shortest time. Specifically, with 90\% accuracy as the target, \xronos reduces time-to-accuracy by 53.6\% and 16.4\% compared with Asteroid and Megatron-LM, respectively. The results show that \xronos effectively reduces fine-tuning time while maintaining the accuracy comparable to existing techniques.

\section{Related work}

\noindent\textbf{Collaborative fine-tuning.}
PipeDream \cite{narayanan2019pipedream} is an early PP-based technique on GPUs without heterogeneity-aware planning. Asteroid \cite{ye2024asteroid} and PAC \cite{ouyang2024pluto} both profile devices and incorporate heterogeneity-aware planning under PP using the same methodology. As PAC shares a similar design with Asteroid and lacks an open-source implementation, we use Asteroid as the representative baseline in our experiments.

DGPAS \cite{li2025dgpas} further extends PP-based training to hybrid GPU+CPU devices with heterogeneity-aware scheduling. 
Al Maruf et al. \cite{al2024optimizing} studies pipelined execution on CPUs but focuses on exploiting multiple cores within a single device rather than coordinating training across multiple edge devices, still using PP. 
Overall, existing techniques are dominated by PP and largely rely on overlapping computation and communication. 
In contrast, \xronos targets collaborative fine-tuning on heterogeneous CPU devices and leverages TP with heterogeneity-aware planning.

\noindent\textbf{TP.}
Several studies have improved TP, largely developed for training on datacenter GPUs. 
Their primary goal is to improve scalability by sharding large tensors and optimizing communication. 
Megatron-LM~\cite{shoeybi2019megatron}, Optimus~\cite{xu2023efficient}, and TAPAS~\cite{shi2025tapas} are representative techniques. 
As such systems are typically composed of similar devices, uniform sharding is commonly used in practice. 
Thus, these techniques are not designed for heterogeneous CPU-only edge devices, which can lead to high idle time, as we demonstrate. 
In contrast, \xronos targets heterogeneous CPUs and allocates tensor partitions adaptively to device capacity, which differs from existing TP techniques.

\section{Conclusion}
This study proposes \xronos, a collaborative fine-tuning system for LLMs on heterogeneous CPU-based edge devices. Through system analysis, we identify that pipeline parallelism suffers from CPU resource contention, while existing tensor parallelism frameworks fail to account for device heterogeneity. To address these challenges, \xronos adopts tensor parallelism with a heterogeneity-aware partitioner that assigns non-uniform workloads across devices. Our evaluations demonstrate that \xronos reduces iteration time $\sim$56\% over Asteroid and reduces idle time $\sim$5.9$\times$ over Megatron-LM.

\section*{Acknowledgment}
This research was supported by Basic Science Research Program through National Research Foundation of Korea (NRF), funded by Ministry of Education (MOE) (RS-2021-NR060143), by NRF grant funded by Korea government (MSIT) (RS-2024-00336564), by IITP-ICT Creative Consilience Program grant funded by MSIT (IITP-2026-RS-2020-II201819), by IITP grant funded by MSIT (RS-2026-25518394), and by ANCHOR program through Seoul ANCHOR Center, funded by MOE and Seoul Metropolitan Government (2026-ANCHOR-01-003-09). Corresponding authors: Gyeongsik Yang and Chuck Yoo.

\bibliographystyle{IEEEtran}
\bibliography{reference_short}

\end{document}